\documentstyle[12pt]{article}

\advance \topmargin by -\headheight
\advance \topmargin by -\headsep
     
\evensidemargin \oddsidemargin
 \date{}    
\begin{document}
\newcommand{\sect}[1]{\setcounter{equation}{0}\section{#1}}
\renewcommand{\theequation}{\thesection.\arabic{equation}}

\topmargin -.6in
\def\nonu{\nonumber}
\def\rf#1{(\ref{eq:#1})}
\def\lab#1{\label{eq:#1}} 
\def\br{\begin{eqnarray}}
\def\er{\end{eqnarray}}
\def\be{\begin{equation}}
\def\ee{\end{equation}}
\def\0{\nonumber}
\def\lb{\lbrack}
\def\rb{\rbrack}
\def\({\left(}
\def\){\right)}
\def\v{\vert}
\def\bv{\bigm\vert}
\def\lskip{\vskip\baselineskip\vskip-\parskip\noindent}
\relax
\newcommand{\nit}{\noindent}
\newcommand{\ct}[1]{\cite{#1}}
\newcommand{\bi}[1]{\bibitem{#1}}
\def\a{\alpha}
\def\b{\beta}
\def\ca{{\cal A}}
\def\cm{{\cal M}}
\def\cn{{\cal N}}
\def\cf{{\cal F}}
\def\d{\delta} 
\def\D{\Delta}
\def\eps{\epsilon}
\def\g{\gamma}
\def\G{\Gamma}
\def\grad{\nabla}
\def\h{ {1\over 2}  }
\def\hc{\hat{c}}
\def\hd{\hat{d}}
\def\hg{\hat{g}}
\def\hp{ {+{1\over 2}}  }
\def\hm{ {-{1\over 2}}  }
\def\k{\kappa}
\def\l{\lambda}
\def\L{\Lambda}
\def\lg{\langle}
\def\m{\mu}
\def\n{\nu}
\def\o{\over}
\def\om{\omega}
\def\O{\Omega}
\def\p{\phi}
\def\pa{\partial}
\def\pr{\prime}
\def\ra{\rightarrow}
\def\rh{\rho}
\def\rg{\rangle}
\def\s{\sigma}
\def\t{\tau}
\def\th{\theta}
\def\ti{\tilde}
\def\wti{\widetilde}
\def\inte{\int dx }
\def\xb{\bar{x}}
\def\yb{\bar{y}}

\def\tr{\mathop{\rm tr}}
\def\Tr{\mathop{\rm Tr}}
\def\partder#1#2{{\partial #1\over\partial #2}}
\def\ds{{\cal D}_s}
\def\wtwo{{\wti W}_2}
\def\lie{{\cal G}}
\def\alie{{\widehat \lie}}
\def\dlie{{\cal G}^{\ast}}
\def\elie{{\widetilde \lie}}
\def\edlie{{\elie}^{\ast}}
\def\hlie{{\cal H}}
\def\wlie{{\widetilde \lie}}

\def\rlx{\relax\leavevmode}
\def\inbar{\vrule height1.5ex width.4pt depth0pt}
\def\IZ{\rlx\hbox{\sf Z\kern-.4em Z}}
\def\IR{\rlx\hbox{\rm I\kern-.18em R}}
\def\IC{\rlx\hbox{\,$\inbar\kern-.3em{\rm C}$}}
\def\one{\hbox{{1}\kern-.25em\hbox{l}}}

\def\PRL#1#2#3{{\sl Phys. Rev. Lett.} {\bf#1} (#2) #3}
\def\NPB#1#2#3{{\sl Nucl. Phys.} {\bf B#1} (#2) #3}
\def\NPBFS#1#2#3#4{{\sl Nucl. Phys.} {\bf B#2} [FS#1] (#3) #4}
\def\CMP#1#2#3{{\sl Commun. Math. Phys.} {\bf #1} (#2) #3}
\def\PRD#1#2#3{{\sl Phys. Rev.} {\bf D#1} (#2) #3}
\def\PLA#1#2#3{{\sl Phys. Lett.} {\bf #1A} (#2) #3}
\def\PLB#1#2#3{{\sl Phys. Lett.} {\bf #1B} (#2) #3}
\def\JMP#1#2#3{{\sl J. Math. Phys.} {\bf #1} (#2) #3}
\def\PTP#1#2#3{{\sl Prog. Theor. Phys.} {\bf #1} (#2) #3}
\def\SPTP#1#2#3{{\sl Suppl. Prog. Theor. Phys.} {\bf #1} (#2) #3}
\def\AoP#1#2#3{{\sl Ann. of Phys.} {\bf #1} (#2) #3}
\def\PNAS#1#2#3{{\sl Proc. Natl. Acad. Sci. USA} {\bf #1} (#2) #3}
\def\RMP#1#2#3{{\sl Rev. Mod. Phys.} {\bf #1} (#2) #3}
\def\PR#1#2#3{{\sl Phys. Reports} {\bf #1} (#2) #3}
\def\AoM#1#2#3{{\sl Ann. of Math.} {\bf #1} (#2) #3}
\def\UMN#1#2#3{{\sl Usp. Mat. Nauk} {\bf #1} (#2) #3}
\def\FAP#1#2#3{{\sl Funkt. Anal. Prilozheniya} {\bf #1} (#2) #3}
\def\FAaIA#1#2#3{{\sl Functional Analysis and Its Application} {\bf #1} (#2)
#3}
\def\BAMS#1#2#3{{\sl Bull. Am. Math. Soc.} {\bf #1} (#2) #3}
\def\TAMS#1#2#3{{\sl Trans. Am. Math. Soc.} {\bf #1} (#2) #3}
\def\InvM#1#2#3{{\sl Invent. Math.} {\bf #1} (#2) #3}
\def\LMP#1#2#3{{\sl Letters in Math. Phys.} {\bf #1} (#2) #3}
\def\IJMPA#1#2#3{{\sl Int. J. Mod. Phys.} {\bf A#1} (#2) #3}
\def\AdM#1#2#3{{\sl Advances in Math.} {\bf #1} (#2) #3}
\def\RMaP#1#2#3{{\sl Reports on Math. Phys.} {\bf #1} (#2) #3}
\def\IJM#1#2#3{{\sl Ill. J. Math.} {\bf #1} (#2) #3}
\def\APP#1#2#3{{\sl Acta Phys. Polon.} {\bf #1} (#2) #3}
\def\TMP#1#2#3{{\sl Theor. Mat. Phys.} {\bf #1} (#2) #3}
\def\JPA#1#2#3{{\sl J. Physics} {\bf A#1} (#2) #3}
\def\JSM#1#2#3{{\sl J. Soviet Math.} {\bf #1} (#2) #3}
\def\MPLA#1#2#3{{\sl Mod. Phys. Lett.} {\bf A#1} (#2) #3}
\def\JETP#1#2#3{{\sl Sov. Phys. JETP} {\bf #1} (#2) #3}
\def\JETPL#1#2#3{{\sl  Sov. Phys. JETP Lett.} {\bf #1} (#2) #3}
\def\PHSA#1#2#3{{\sl Physica} {\bf A#1} (#2) #3}
\def\PHSD#1#2#3{{\sl Physica} {\bf D#1} (#2) #3}
%new-defects-n2.tex\hskip 8cm  \today
% Ja corrigido Solucao geral 
\begin{center}
{\large\bf    Integrablility of a Classical   $N= 2$ Super Sinh-Gordon Model with Jump Defects }
\end{center}
\normalsize
\vskip .4in

\begin{center}
J.F. Gomes,  L.H. Ymai and A.H. Zimerman

\par \vskip .1in \noindent
Instituto de F\'{\i}sica Te\'{o}rica-UNESP\\
Rua Pamplona 145\\
01405-900 S\~{a}o Paulo, Brazil
\par \vskip .3in

\end{center}

\begin{abstract}
The Lagrangian formalism for the $N=2$ supersymmetric sinh-Gordon model with a jump defect is considered. 
 The modified conserved momentum and energy are constructed in terms of border functions.  The supersymmetric Backlund 
 transformation is given and an one-soliton solution is obtained.  

The Lax formulation based on the affine super Lie algebra $sl(2,2)$ within the space  split by the defect leads to the integrability of the model and 
henceforth to the existence of an infinite number of constants of motion.

\end{abstract}

\section{Introduction}

In reference \cite{mussardo} a quantum  theory of free bosons and free fermions   
subject to   an internal boundary condition (jump defect) preserving the integrability
was considered.  More recently a classical Lagrangian approach   was proposed 
for certain class of non linearly interacting  bosonic fields \cite{bowcock1},  \cite{bowcock2}, \cite{corrigan}. 
 The authors have  considered  a class of systems described 
by internal boundary conditions corresponding to Backlund transformations and showed to 
preserve integrability.  
Such framework was generalized in \cite{ymai}
to include  both bosons and fermions 
interacting  in a non linear manner.  Specifically we have considered  the super 
sinh-Gordon model with $N=1$.  The border functions were constructed and shown to give rise to Backlund 
transformations.   The integrability of the system in the presence of this kind of defect was  considered in terms of zero curvature representation.

The $N=2$ super sinh-Gordon model was  proposed in \cite{ik}, \cite{ku} using superfield formalism.  
Much later, a systematic construction was developed  from the algebraic formalism point of view 
where the $N=2$ super sinh-Gordon equations appears to be a member of an integrable hierarchy \cite{nosso1}, \cite{nosso2}.  
The  description of integrability in terms of an affine algebraic
structure has provided a  neat  universal  framework from which the dynamics, conservation laws,
soliton solutions, etc can be constructed and studied.  Following the same line of reasoning, 
supersymmetry transformation was also incorporated into the  framework (see e.g. \cite{nosso1}, \cite{nosso2}) 
associated to half integer gradations.

The purpose of  this paper is to extend the results obtained in \cite{ymai} for the $N=2$ super sinh-Gordon model.  In sect. 2 we discuss the Lagrangian formalism introducing the jump defect in terms of the border functions.  These are defined by a modified momentum conservation which is consistent with the Backlund transfomation for the $N=2$ super sinh-Gordon.
The presence of the defect  require modification of the conserved  energy of the system.
In section 3 we extend the construction of Backlund transformation of ref. \cite{chai} in terms of super fields to the $N=2$ super sinh-Gordon.  We also write down the the Backlund transformation in components and obtain a one-soliton solution.  
These  are invariant under the $N=2$ supersymmetry transformation.    In section 4 we present the zero curvature  formulation in terms of an affine $sl(2,2)$ super Lie algebra. By introducing two regions around the defect, we explicitly construct, in a closed form, a gauge group element connection the Lax pair in the overlap region.  This fact guarantees the existence of an infinite set of conservation laws.   

\section{Lagrangian Description }

The starting point is the Lagrangian density describing  the $N=2$ super sinh-Gordon theory with bosonic  $\phi_1, \varphi_1$ and fermionic $\bar \psi_1, \psi_1, \bar \chi_1, \chi_1$ fields
in the region $x<0$ and corresponding $\phi_2, \varphi_2$ and $\bar \psi_2, \psi_2, \bar \chi_2, \chi_2$ for $x>0$,
\br
{\cal L}_p &= & {1\o 2} (\pa_x \phi_p)^2 -{1\o 2} (\pa_t \phi_p)^2 + 2\bar \psi_p \pa_t \bar \psi_p + 2\bar \psi_p \pa_x \bar \psi_p  + 2\psi_p \pa_t  \psi_p - 2\psi_p \pa_x  \psi_p \nonu \\
 &-& {1\o 2} (\pa_x \varphi_p)^2 +{1\o 2} (\pa_t \varphi_p)^2 -2 \bar
\chi_p \pa_t \bar \chi_p - 2\bar \chi_p \pa_x \bar \chi_p
-2 \chi_p \pa_t  \chi_p + 2\chi_p \pa_x  \chi_p \nonu \\
 &-& 16 (\psi_p \bar \psi_p + \chi_p \bar \chi_p) \cosh \varphi_p \cosh \phi_p - 4 \cosh (2 \varphi_p) \nonu \\
 &+& 16 (\psi_p \bar \chi_p + \chi_p \bar \psi_p) \sinh \varphi_p \sinh \phi_p + 4 \cosh (2\phi_p), \quad \quad p=1,2.
\label{1}
\er
We shall now consider the system with a defect  at the origin ($x=0$) described by 
\br
{\cal L} = \theta (-x) {\cal L}_{1} + \theta (x) {\cal L}_{2} + \d(x) {\cal L}_D
\label{3}
\er
where 
\br 
{\cal L}_{D} &=&  {{1/ 2}}(\phi_2 \pa_t \phi_1 - \phi_1 \pa_t
\phi_2) -2\psi_1 \psi_2 - 2\bar \psi_1 \bar \psi_2
+  \zeta_1^- \pa_t  \zeta_1^+ \nonu \\ 
 \nonu
& - &{1/ 2} (\varphi_2 \pa_t \varphi_1 - \varphi_1 \pa_t \varphi_2)
+2\chi_1 \chi_2 + 2\bar \chi_1 \bar \chi_2
+   \zeta_1^+ \pa_t  \zeta_1^- \nonu \\ 
& + & B_0(\phi_p, \varphi_p) + B_1 (\phi_p, \varphi_p, \psi_p, \chi_p, \bar \psi_p, \bar \chi_p,  \zeta_1^-,  \zeta_1^+).
\label{4}
\er
The boundary functions $B= B_0 + B_1$ describe the defect and  $\zeta_1^{\pm}$ are fermionic auxiliary fields.  Notice that they describe the effect of the jump defect on the fermionic fields generalizing the $N=1$ case \cite{ymai} where one auxiliary field appears.

The equations of motion are
\br
(\partial_x^{2}-\partial_{t}^{2})\phi_p &=&
8\mathrm{sinh}(2\phi_p)-16(\psi_p \bar{\psi}_p + \chi_p
\bar{\chi}_p)\mathrm{sinh}\phi_p
\mathrm{cosh}\varphi_p\nonumber\\
&&+16(\psi_p \bar{\chi}_p + \chi_p \bar{\psi}_p)\mathrm{sinh}\varphi_p
\mathrm{cosh}\phi_p\nonumber\\
(\partial_x^{2}-\partial_{t}^{2})\varphi_p &=&
8\mathrm{sinh}(2\varphi_p)+16(\psi_p \bar{\psi}_p + \chi_p
\bar{\chi}_p)\mathrm{sinh}\varphi_p
\mathrm{cosh}\phi_p\nonumber\\
&&-16(\psi_p \bar{\chi}_p + \chi_p \bar{\psi}_p)\mathrm{sinh}\phi_p
\mathrm{cosh}\varphi_p\nonumber\\
(\partial_x - \partial_t)\psi_p &=& -4\bar{\psi}_p
\mathrm{cosh}\phi_p\mathrm{cosh}\varphi_p + 4\bar{\chi}_p
\mathrm{sinh}\phi_p\mathrm{sinh}\varphi_p\nonumber\\
(\partial_x - \partial_t)\chi_p &=& -4\bar{\psi}_p
\mathrm{sinh}\phi_p\mathrm{sinh}\varphi_p + 4\bar{\chi}_p
\mathrm{cosh}\phi_p\mathrm{cosh}\varphi_p\nonumber\\
(\partial_x + \partial_t)\bar{\psi}_p &=& -4\psi_p
\mathrm{cosh}\phi_p\mathrm{cosh}\varphi_p + 4\chi_p
\mathrm{sinh}\phi_p\mathrm{sinh}\varphi_p\nonumber\\
(\partial_x + \partial_t)\bar{\chi}_p &=& -4\psi_p
\mathrm{sinh}\phi_p\mathrm{sinh}\varphi_p + 4\chi_p
\mathrm{cosh}\phi_p\mathrm{cosh}\varphi_p
\label{5}
\er
where  $p=1,2$ corresponds to $x<0$ or $x>0$ respectively.  These equations are invariant under the supersymmetry transformation,
\br
\d (\phi_p \pm \varphi_p) = 2 (\psi_p \mp \chi_p) \eps_{\pm},  \quad \d (\psi_p \pm \chi_p) = - 1/2(\pa_x + \pa_t) (\phi_p \mp \varphi_p) \eps_{\pm},\nonu \\ 
\d  (\bar \psi_p \pm \bar \chi_p) = 2 \sinh (\phi_p \pm \varphi_p) \eps_{\mp}.
 \label{supe}
\er
where $\eps_{\pm}$ are fermionic parameters.
In $x=0$ we have, 
\br
& \pa_x \phi_1 - \pa_t \phi_2 =  - \pa_{\phi_1} B,   &\quad \quad 
\pa_x \phi_2 - \pa_t \phi_1 =   \pa_{\phi_2} B, \nonu \\
& \pa_x \varphi_1 - \pa_t \varphi_2 =   \pa_{\varphi_1} B,  &\quad \quad
\pa_x \varphi_2 - \pa_t \varphi_1 =   -\pa_{\varphi_2} B, \nonu \\
& \psi_1 -\psi_2 = -{1\o 2}\pa_{\psi_1} B=-{1\o 2}\pa_{\psi_2} B   & \quad \quad
\chi_1 -\chi_2 = {1\o 2}\pa_{\chi_1} B={1\o 2}\pa_{\chi_2} B\nonu \\
& \bar \psi_1 + \bar \psi_2 = {1\o 2}\pa_{\bar \psi_1}B=-{1\o 2}\pa_{\bar \psi_2}B,  & \quad \quad 
\bar \chi_1 + \bar \chi_2 = -{1\o 2}\pa_{\bar \chi_1}B={1\o 2}\pa_{\bar \chi_2}B, \nonu \\
& { \pa_t \zeta_1^-} = -{1\o 2} \pa_{ \zeta_1^+} B,  & \quad \quad 
{\pa_t  \zeta_1^+} = -{1\o 2} \pa_{ \zeta_1^-} B
\label{6}
\er
%which can also be shown to be invariant under (\ref{sup}).
%which are invariant under $N=2$ supersyummetry transformation, i.e.,
%\br
%\d (\phi_p \pm \varphi_p) = 2 (\psi_p \pm \chi_p) \eps_{\pm},  \quad \d (\psi_p \pm \chi_p) = - \pa_x (\phi_p \pm \varphi_p) \eps_{\pm},\quad 
%\d  (\bar \psi_p \pm \bar \chi_p) = 2 \sinh (\phi_p \pm \varphi_p) \eps_{\pm}.
% \label{2}
%\nonu
%\er

The canonical momentum $P$,   given by
\br
P &=& \int_{-\infty}^0 \( \pa_x \phi_1 \pa_t \phi_1 -2 \bar \psi_1 \pa_x \bar \psi_1 -2 \psi_1 \pa_x \psi_1 \right. 
- \left. \pa_x \varphi_1 \pa_t \varphi_1 + 2 \bar \chi_1 \pa_x \bar \chi_1 + 2 \chi_1 \pa_x \chi_1\) dx 
\nonu \\
&+&
 \int^{\infty}_0 \( \pa_x \phi_2 \pa_t \phi_2 -2 \bar \psi_2 \pa_x \bar \psi_2 -2 \psi_2 \pa_x \psi_2 \right.
-\left. \pa_x \varphi_2 \pa_t \varphi_2 +2 \bar \chi_2 \pa_x \bar \chi_2 + 2\chi_2 \pa_x \chi_2\) dx 
\nonu \\
 \label{8} 
 \er
is not conserved in time due to the presence of the defect.  Instead we find after making use of the equations of motion,
\br
\dot{P}&=&\Big[\frac{1}{2}(\partial_t
\phi_1)^2-\frac{1}{2}(\partial_t
\varphi_1)^2+\frac{1}{2}(\partial_x\phi_1)^2-\frac{1}{2}(\partial_x\varphi_1)^2
-2\bar{\psi}_1 \partial_t \bar{\psi}_1 - 2\psi_1 \partial_t \psi_1
\nonumber\\
&&+ 2\bar{\chi}_1 \partial_t \bar{\chi}_1 + 2\chi_1
\partial_t\chi_1
-4\mathrm{cosh}(2\phi_1)+4\mathrm{cosh}(2\varphi_1)+16(\psi_1
\bar{\psi}_1 + \chi_1 \bar{\chi}_1)\mathrm{cosh}\phi_1
\mathrm{cosh}\varphi_1\nonumber\\
&&-16(\psi_1 \bar{\chi}_1 + \chi_1 \bar{\psi}_1)\mathrm{sinh}\phi_1
\mathrm{sinh}\varphi_1\Big]_{x=0} - \Big[(1\to 2)\Big]_{x=0} \nonumber
\er
Factorizing $B_0 = B_{0}^{(+)}+B_{0}^{(-)}$  and $B_1 = B_{1}^{(+)}+B_{1}^{(-)}$, where $ \phi_{\pm} = \phi_1 \pm \phi_2,
\quad  \varphi_{\pm} = \varphi_1 \pm \varphi_2, \cdots $ such that
\br
B_{0}^{(+)}&=&  B_{0}^{(+)}(\phi_{+},\varphi_{+}), \quad   \quad  
 B_{0}^{(-)}= B_{0}^{(-)}(\phi_{-},\varphi_{-}) \nonu \\
 B_{1}^{(+)}&=&  B_{1}^{(+)}(\phi_{+},\varphi_{+},\psi_{+},\chi_{+}, \zeta_1^-,  \zeta_1^+), \quad  \quad  
 B_{1}^{(-)}= B_{1}^{(-)}(\phi_{-},\varphi_{-},\bar{\psi}_{-},\bar{\chi}_{-}, \zeta_1^-,  \zeta_1^+) 
 \label{10}
 \er
 and using eqns. (\ref{6}), we define the modified momentum 
 \br
 {\cal P} = P + \Big[B_0^{(+)} - B_0^{(-)}+ B_1^{(+)} - B_1^{(-)} 
+ 2\bar \psi_1 \bar \psi_2 -2\psi_1 \psi_2 - 2\bar \chi_1 \bar
\chi_2 + 2\chi_1 \chi_2 \Big]|_{x=0}
\label{10a}
 \er 
which is conserved in time provided
the border functions $B_0$ and $B_1$  satisfy
\br
  \pa_{\phi_+} B_0^{(+)} \pa_{\phi_-}B_0^{(-)} -  \pa_{\varphi_+} B_0^{(+)} \pa_{\varphi_-}B_0^{(-)}  
 = 4 \sinh \phi_+ \sinh \phi_- - 4 \sinh \varphi_+ \sinh \varphi_-
\label{11}
\er
and 
\br
& & \pa_{\phi_+} B_0^{(+)} \pa_{\phi_-} B_1^{(-)} + \pa_{\phi_-} B_0^{(-)} \pa_{\phi_+} B_1^{(+)}  
- \pa_{\varphi_+} B_0^{(+)} \pa_{\varphi_-} B_1^{(-)} - \pa_{\varphi_-} B_0^{(-)} \pa_{\varphi_+} B_1^{(+)} \nonu \\
&+&\pa_{\phi_+} B_1^{(+)} \pa_{\phi_-} B_1^{(-)} - \pa_{\varphi_+} B_1^{(+)} \pa_{\varphi_-} B_1^{(-)}  
- {1\o 2} (\pa_{ \zeta_1^-} B_1^{(-)} \pa_{ \zeta_1^+}B_1^{(+)} + \pa_{ \zeta_1^+} B_1^{(-)} \pa_{ \zeta_1^-}B_1^{(+)}) \nonu \\
&=&
-2(\psi_+ \bar \psi_+ + \psi_- \bar \psi_- +\chi_+ \bar \chi_+ + \chi_- \bar \chi_-)\L_- 
- 2(\psi_+ \bar \psi_- + \psi_- \bar \psi_+ +\chi_+ \bar \chi_- + \chi_- \bar \chi_+)\L_+ \nonu \\
&+& 2(\psi_+ \bar \chi_+ + \psi_- \bar \chi_- +\chi_+ \bar \psi_+ + \chi_- \bar \psi_-)\D_- 
+ 2(\psi_+ \bar \chi_- + \psi_- \bar \chi_+ +\chi_+ \bar \psi_- + \chi_- \bar \psi_+)\D_+ \nonu \\
\label{12}
\er
where
\br
\L_{\pm} &=& \cosh ({{\phi_+ + \phi_-}\o 2}) \cosh ({{\varphi_+ + \varphi_-}\o 2}) \pm \cosh ({{\phi_+ - \phi_-}\o 2}) \cosh ({{\varphi_+ - \varphi_-}\o 2}), \nonu \\
\D_{\pm} &=& \sinh ({{\phi_+ + \phi_-}\o 2}) \sinh ({{\varphi_+ + \varphi_-}\o 2}) \pm \sinh ({{\phi_+ - \phi_-}\o 2}) \sinh ({{\varphi_+ - \varphi_-}\o 2}), \label{13}
\er
%%%%%%%%%%%%%%%%%%%%%%%%%%%%%%%%%%%%%%%%%%%%%%%%%%%%%%%%%%%%%%%%%%%%%%%%%%%%%%%%%%%%%%%%%%%%%%%%%%%%%%%%%%%%%%%%%%%%%%
The energy of the system with the defect is given by
\begin{eqnarray}
E=\int_{-\infty}^{0}dx\,{\mathcal{H}}_1+\int_{0}^{\infty}dx\,{\mathcal{H}}_2,\nonumber
\end{eqnarray}
where
\begin{eqnarray}
{\mathcal{H}}_p&=&\left[\frac{1}{2}(\partial_x\phi_p)^2+\frac{1}{2}(\partial_t\phi_p)^2-\frac{1}{2}(\partial_x\varphi_p)^2-\frac{1}{2}(\partial_t\varphi_p)^2-2\psi_p\partial_x\psi_p+2\bar{\psi}_p\partial_x\bar{\psi}_p\right.\nonumber\\
&+&2\chi_p\partial_x\chi_p-2\bar{\chi}_p\partial_x\bar{\chi}_p-16(\psi_p\bar{\psi}_p+\chi_p\bar{\chi}_p)\mathrm{cosh}\varphi_p
\,\mathrm{cosh}\phi_p \nonu \\
&+&16(\psi_p\bar{\chi}_p+\chi_p\bar{\psi}_p)\mathrm{sinh}\varphi_p
\,\mathrm{sinh}\phi_p
+4\mathrm{cosh}(2\phi_p)-4\mathrm{cosh}(2\varphi_p)\Bigg], \quad p=1,2 
\end{eqnarray}
It follows after using the equations of motion (\ref{5}) that 
\begin{eqnarray}
\frac{dE}{dt}&=&\left[\partial_x\phi_1\partial_t\phi_1-\partial_x\varphi_1\partial_t\varphi_1-2\psi_1\partial_t\psi_1+2\bar{\psi}_1\partial_t\bar{\psi}_1+2\chi_1\partial_t\chi_1-2\bar{\chi}_1\partial_t\bar{\chi}_1\right]_{x=0}\nonumber\\
&&-\left[\partial_x\phi_2\partial_t\phi_2-\partial_x\varphi_2\partial_t\varphi_2-2\psi_2\partial_t\psi_2+2\bar{\psi}_2\partial_t\bar{\psi}_2+2\chi_2\partial_t\chi_2-2\bar{\chi}_2\partial_t\bar{\chi}_2\right]_{x=0}.\nonumber
\end{eqnarray}
Inserting the Backlund transformation (\ref{6})  we can define the modified conserved energy,
\begin{eqnarray}
{\mathcal{E}} =E+ [B-2\psi_1\psi_2+2\chi_1\chi_2-2\bar{\psi}_1\bar{\psi}_2+2\bar{\chi}_1\bar{\chi}_2]|_{x=0},
\end{eqnarray}
where $B = B_0^{(+)} + B_0^{(-)}+ B_1^{(+)} + B_1^{(-)}$.

%%%%%%%%%%%%%%%%%%%%%%%%%%%%%%%%%%%%%%%%%%%%%%%%%%%%%%%%%%%%%%%%%%%%%%%%%%%%%%%%%%%%%%%%%%%%%%%%%%%%%%%%%%%%%%%%%%%%%%
Equation (\ref{11}) has the following solution
\br
B_0^{(+)}  &=& B_0^{(+)} (\phi_+, \varphi_+) = {{2 \b_3}\o {\b_2}}\( \cosh \phi_+ -  \cosh \varphi_+\), \nonu \\
B_0^{(-)}  &=& B_0^{(-)} (\phi_-, \varphi_-) = {{2 \b_2}\o {\b_3}} \(\cosh \phi_- -  \cosh \varphi_-\). 
\label{15}
\er
The solution of (\ref{12}) is verified  for 
\br
B_1^{(+)}  &=&  {{i}\o {\sqrt{2}}} \zeta_1^-\(-\b_3 (\psi_+ - \chi_+)\cosh {1\o 2} (\phi_+ + \varphi_+) \) \nonu \\
  &+&  {{i}\o {\sqrt{2}}} \zeta_1^+ {{\b_1}\o {\b_3}}\( {-\b_3} (\psi_+ + \chi_+)\cosh {1\o 2} (\phi_+ - \varphi_+)\), \label{2.15}  
 \er
 \br
B_1^{(-)}  &=&  {{i}\o {\sqrt{2}}}\zeta_1^-\(\b_2 (\bar \psi_- + \bar \chi_-)\cosh {1\o 2} (\phi_- - \varphi_-) \) \nonu \\
  &+&  {{i}\o {\sqrt{2}}} \zeta_1^+ {{\b_1}\o {\b_3}}\( {\b_2} (\bar \psi_- - \bar \chi_-)\cosh {1\o 2} (\phi_- + \varphi_-) \) 
  \label{16}
\er
where $\b_1, \b_2$ and $\b_3$ are arbitrary constants.

The solution given above also verify the following  identities 

\begin{eqnarray}\label{expr1}
&&(\partial_{\phi_{+}}B_{0}^{(+)}\partial_{\phi_{-}}B_{1}^{(-)}+\partial_{\phi_{-}}B_{0}^{(-)}\partial_{\phi_{+}}B_{1}^{(+)})
-(\partial_{\varphi_{+}}B_{0}^{(+)}\partial_{\varphi_{-}}B_{1}^{(-)}+\partial_{\varphi_{-}}B_{0}^{(-)}\partial_{\varphi_{+}}B_{1}^{(+)})\nonumber\\
&=&-2(\psi_{+}\bar{\psi}_{+}+\psi_{-}\bar{\psi}_{-}+\chi_{+}\bar{\chi}_{+}+\chi_{-}\bar{\chi}_{-})\Lambda_{-}
+2(\psi_{+}\bar{\chi}_{+}+\psi_{-}\bar{\chi}_{-}+\chi_{+}\bar{\psi}_{+}+\chi_{-}\bar{\psi}_{-})\Delta_{-}
\nonu \\
\end{eqnarray}

\begin{eqnarray}\label{expr3}
\frac{1}{2}(\partial_{\zeta_{1}^{-}}B_{1}^{(-)}\partial_{\zeta_{1}^{+}}B_{1}^{(+)}+\partial_{\zeta_{1}^{+}}B_{1}^{(-)}\partial_{\zeta_{1}^{-}}B_{1}^{(+)})
=2(\psi_{+}\bar{\psi}_{-}+\chi_{+}\bar{\chi}_{-})\Lambda_{+}-2(\psi_{+}\bar{\chi}_{-}+\chi_{+}\bar{\psi}_{-})\Delta_{+}
\end{eqnarray}

\begin{eqnarray}\label{expr2}
(\partial_{\phi_{+}}B_{1}^{(+)}\partial_{\phi_{-}}B_{1}^{(-)}-\partial_{\varphi_{+}}B_{1}^{(+)}\partial_{\varphi_{-}}B_{1}^{(-)})=0
\end{eqnarray}

\br
(\psi_- \bar \psi_+ + \chi_- \bar \chi_+)\L_+ - (\psi_- \bar \chi_+ + \chi_- \bar \psi_+)\Delta_+ = 0
\label{del}
\er
In analogy with ref. \cite{ymai} where we have dealt with the $N=1$ case, the space derivatives of $\zeta_1^{\pm}$, i.e. $\pa_x \zeta_1^{\pm}$ can be obtained  by requiring compatibility of eqns. (\ref{6}) with (\ref{5}) with $B$ given by (\ref{15})-(\ref{16}).  Explicit expressions are given in (\ref{zet1}) and (\ref{zet2}) in consistency with the Backlund transformation where the auxiliary fields $\zeta_1^{\pm}$ appear in a natural manner.

\section{Backlund Transformation - One-soliton Solution}

Introducing 
\br
\pa_z = {1\o 2}(\pa_x + \pa_t), \quad \pa_{\bar z} = {1\o 2}(\pa_t - \pa_x), \quad \quad   z = x+t, \quad  \bar z = t-x
\er
we define, according to ref. \cite{ku}, the super fields 
\begin{eqnarray}
\Upsilon^{+}&=&\eta^{+}(z^{+},\bar{z}^{+})+\theta^{+}\psi^{-}(z^{+},\bar{z}^{+})+\bar{\theta}^{+}\bar{\psi}^{-}(z^{+},\bar{z}^{+})+\theta^{+}\bar{\theta}^{+}F^{+}(z^{+},\bar{z}^{+})\nonumber\\
\Upsilon^{-}&=&\eta^{-}(z^{-},\bar{z}^{-})+\theta^{-}\psi^{+}(z^{-},\bar{z}^{-})+\bar{\theta}^{-}\bar{\psi}^{+}(z^{-},\bar{z}^{-})+\theta^{-}\bar{\theta}^{-}F^{-}(z^{-},\bar{z}^{-})\nonumber
\end{eqnarray}
where
\begin{eqnarray}
z^{\pm}=z\pm\frac{1}{2}\theta^{+}\theta^{-}\quad \quad 
\bar{z}^{\pm}=\bar{z}\pm\frac{1}{2}\bar{\theta}^{+}\bar{\theta}^{-}\nonumber
\end{eqnarray}
and the super derivatives
\begin{eqnarray}
D_{+}=\frac{\partial}{\partial\theta^{+}}+\frac{1}{2}\theta^{-}\partial_{z}, \qquad
\bar{D}_{+}=\frac{\partial}{\partial\bar{\theta}^{+}}+\frac{1}{2}\bar{\theta}^{-}\partial_{\bar{z}},\nonumber\\
D_{-}=\frac{\partial}{\partial\theta^{-}}+\frac{1}{2}\theta^{+}\partial_{z},
\qquad
\bar{D}_{-}=\frac{\partial}{\partial\bar{\theta}^{-}}+\frac{1}{2}\bar{\theta}^{+}\partial_{\bar{z}},\nonumber
\end{eqnarray}
The equations of motion are then given by \cite{ku}
\begin{eqnarray}
\bar{D}_{+}D_{+}\Upsilon^{+}=g\, {{\mathrm{sin}}}\Upsilon^{-}, \qquad
\bar{D}_{-}D_{-}\Upsilon^{-}=g\, {{\mathrm{sin}}}\Upsilon^{+}
\label{eq}
\end{eqnarray}
where $g$ is a constant.  In  components we find  for the first equation (\ref{eq})
\begin{eqnarray}
F^{+}&=&g\, {\mathrm{sin}}\eta^{-}\quad \quad 
\partial_{\bar{z}}\psi^{-}=g\,{\mathrm{cos}}\eta^{-}\bar{\psi}^{+}\nonumber\\
\partial_{z}\bar{\psi}^{-}&=&-g\,{\mathrm{cos}}\eta^{-}\psi^{+}\quad \quad 
\partial_{z}\partial_{\bar{z}}\eta^{+}=-g\,{\mathrm{cos}}\eta^{-}F^{-}-g\,{\mathrm{sin}}\eta^{-}\psi^{+}\bar{\psi}^{+}\nonumber
\end{eqnarray}
while for the second equation (\ref{eq}) we have 
\begin{eqnarray}
F^{-}&=&g\, {\mathrm{sin}}\eta^{+}\quad \quad 
\partial_{\bar{z}}\psi^{+}=g\,{\mathrm{cos}}\eta^{+}\bar{\psi}^{-}\nonumber\\
\partial_{z}\bar{\psi}^{+}&=&-g\,{\mathrm{cos}}\eta^{+}\psi^{-}\quad \quad 
\partial_{z}\partial_{\bar{z}}\eta^{-}=-g\,{\mathrm{cos}}\eta^{+}F^{+}-g\,{\mathrm{sin}}\eta^{+}\psi^{-}\bar{\psi}^{-}\nonumber
\end{eqnarray}
The superfields $\Upsilon^{+}$ e $\Upsilon^{-}$ are chiral and satisfy
\begin{eqnarray}
\bar{D}_{+}\Upsilon^{-}=D_{+}\Upsilon^{-}=0, \qquad
\bar{D}_{-}\Upsilon^{+}=D_{-}\Upsilon^{+}=0
\label{sup}
\end{eqnarray}

Extending the procedure given in \cite{chai}, we propose for the first eqn. (\ref{eq}) the following Backund transformation
\begin{eqnarray}
D_{+}\Upsilon_{1}^{+}&=&D_{+}\Upsilon_{2}^{+}+\b_{1}{\mathcal{F}}_{1}{\mathrm{cos}}\left(\frac{\Upsilon_{1}^{-}+\Upsilon_{2}^{-}}{2}\right)\label{f1}\\
\bar{D}_{+}\Upsilon_{1}^{+}&=&-\bar{D}_{+}\Upsilon_{2}^{+}+\b_{2}{\mathcal{F}}_{2}{\mathrm{cos}}\left(\frac{\Upsilon_{1}^{-}-\Upsilon_{2}^{-}}{2}\right)\label{f2}
\end{eqnarray}
where  ${\mathcal{F}}_{1}$ and  ${\mathcal{F}}_{2}$ are auxiliary fermionic superfields,
$\b_1$ and  $\b_{2}$ are arbitrary constants. From 
\begin{eqnarray}
(\bar{D}_{+}D_{+}+D_{+}\bar{D}_{+})\Upsilon_{1}^{+}=0
\end{eqnarray}
we obtain 
\begin{eqnarray}
\bar{D}_{+}D_{+}\Upsilon_{2}^{+}=g\, {\mathrm{sin}}\Upsilon_{2}^{-}
\end{eqnarray}
if the auxiliary superfields  ${\mathcal{F}}_{1}$ and  ${\mathcal{F}}_{2}$
satisfy
\begin{eqnarray}
\bar{D}_{+}{\mathcal{F}}_{1}=\frac{2g}{\b_{1}}{\mathrm{sin}}\left(\frac{\Upsilon_{1}^{-}-\Upsilon_{2}^{-}}{2}\right)\quad \quad 
D_{+}{\mathcal{F}}_{2}=-\frac{2g}{\b_{2}}{\mathrm{sin}}\left(\frac{\Upsilon_{1}^{-}+\Upsilon_{2}^{-}}{2}\right)
\end{eqnarray}
For the second equation (\ref{eq}) consider the following transformations
\begin{eqnarray}
D_{-}\Upsilon_{1}^{-}&=&D_{-}\Upsilon_{2}^{-}+\b_{3}{\mathcal{G}}_{1}{\mathrm{cos}}\left(\frac{\Upsilon_{1}^{+}+\Upsilon_{2}^{+}}{2}\right)\label{g1}\\
\bar{D}_{-}\Upsilon_{1}^{-}&=&-\bar{D}_{-}\Upsilon_{2}^{-}+\b_{4}{\mathcal{G}}_{2}{\mathrm{cos}}\left(\frac{\Upsilon_{1}^{+}-\Upsilon_{2}^{+}}{2}\right)\label{g2}
\end{eqnarray}
where ${\mathcal{G}}_{1}$ and ${\mathcal{G}}_{2}$ are auxiliary fermionic superfields
and $\b_3$ and  $\b_{4}$ are arbitrary constants. Similarly, 
from 
\begin{eqnarray}
(\bar{D}_{-}D_{-}+D_{-}\bar{D}_{-})\Upsilon_{1}^{-}=0
\end{eqnarray}
we obtain the second equation
\begin{eqnarray}
\bar{D}_{-}D_{-}\Upsilon_{2}^{-}=g\, {\mathrm{sin}}\Upsilon_{2}^{+}\nonumber
\end{eqnarray}
provided  ${\mathcal{G}}_{1}$ and  ${\mathcal{G}}_{2}$ satisfy the following conditions
\begin{eqnarray}
\bar{D}_{-}{\mathcal{G}}_{1}=\frac{2g}{\b_{3}}{\mathrm{sin}}\left(\frac{\Upsilon_{1}^{+}-\Upsilon_{2}^{+}}{2}\right)\quad \quad 
D_{-}{\mathcal{G}}_{2}=-\frac{2g}{\b_4}{\mathrm{sin}}\left(\frac{\Upsilon_{1}^{+}+\Upsilon_{2}^{+}}{2}\right)
\end{eqnarray}

In the Appendix A we derive further compatibility relations involving the Fermionic auxiliary superfields.  These are written explicitly  in components and provide  algebraic relations like  (\ref{beta}) also.

Choosing $g=2$, redefining fields
\br
\eta^{\pm}_p \rightarrow i (\phi_p \pm \varphi_p), 
\quad \psi_p^{\pm} \rightarrow i\sqrt{2} (\psi_p \pm \chi_p),
 \quad  \bar \psi_p^{\pm} \rightarrow i\sqrt{2} ( \bar \psi_p {\pm} \bar \chi_p), \quad p=1,2.
 \er
and denoting 
\br
\Phi^{(\pm )}_s = \phi_s \pm \varphi_s, \quad  \Psi^{(\pm )}_s = \psi_s \pm \chi_s, \quad \bar {\Psi}^{(\pm )}_s = \bar \psi_s \pm \bar \chi_s, \quad s = \pm
\er
where $\phi_{\pm} = \phi_1 \pm \phi_2,
\quad  \varphi_{\pm} = \varphi_1 \pm \varphi_2, \quad \psi_{\pm} = \psi_1 \pm \psi_2,
\quad  \chi_{\pm} = \chi_1 \pm \chi_2, \cdots$
In components the Backlund transformation reads
\begin{eqnarray}
\partial_x \phi_1 - \partial_t \phi_2&=&\frac{i}{2\sqrt{2}}\zeta_{1}^{-}\left[-\b_{2}\,
\mathrm{sinh}\left(\frac{\Phi_{-}^{(-)}}{2}\right)\bar{\Psi}_{-}^{(+)}+\b_{3}\,\mathrm{sinh}\left(\frac{\Phi_{+}^{(+)}}{2}\right)\Psi_{+}^{(-)}\right]\nonumber\\
&&+\frac{i}{2\sqrt{2}}\frac{\b_1}{\b_3}\zeta_{1}^{+}\left[-\b_{2}\,
\mathrm{sinh}\left(\frac{\Phi_{-}^{(+)}}{2}\right)\bar{\Psi}_{-}^{(-)}+\b_{3}\,\mathrm{sinh}\left(\frac{\Phi_{+}^{(-)}}{2}\right)\Psi_{+}^{(+)}\right]\nonumber\\
&&-\frac{2\b_2}{\b_3}\mathrm{sinh}(\phi_{1}-\phi_{2})-\frac{2\b_3}{\b_2}\mathrm{sinh}(\phi_{1}+\phi_{2})
\end{eqnarray}

\begin{eqnarray}
\partial_x \phi_2 - \partial_t \phi_1&=&\frac{i}{2\sqrt{2}}\zeta_{1}^{-}\left[-\b_{2} \,
\mathrm{sinh}\left(\frac{\Phi_{-}^{(-)}}{2}\right)\bar{\Psi}_{-}^{(+)}-\b_{3}\,\mathrm{sinh}\left(\frac{\Phi_{+}^{(+)}}{2}\right)\Psi_{+}^{(-)}\right]
\nonu \\
&&+\frac{i}{2\sqrt{2}}\frac{\b_1}{\b_3}\zeta_{1}^{+}\left[-\b_{2}\,
\mathrm{sinh}\left(\frac{\Phi_{-}^{(+)}}{2}\right)\bar{\Psi}_{-}^{(-)}-\b_{3}\,\mathrm{sinh}\left(\frac{\Phi_{+}^{(-)}}{2}\right)\Psi_{+}^{(+)}\right]\nonumber\\
&&-\frac{2\b_2}{\b_3}\mathrm{sinh}(\phi_{1}-\phi_{2})+\frac{2\b_3}{\b_2}\mathrm{sinh}(\phi_{1}+\phi_{2})
\end{eqnarray}

\begin{eqnarray}
\partial_x \varphi_1 - \partial_t \varphi_2&=&\frac{i}{2\sqrt{2}}\zeta_{1}^{-}\left[-\b_{2}\,
\mathrm{sinh}\left(\frac{\Phi_{-}^{(-)}}{2}\right)\bar{\Psi}_{-}^{(+)}-\b_{3}\,\mathrm{sinh}\left(\frac{\Phi_{+}^{(+)}}{2}\right)\Psi_{+}^{(-)}\right]\nonumber\\
&&+\frac{i}{2\sqrt{2}}\frac{\b_1}{\b_3}\zeta_{1}^{+}\left[\b_{2}\,
\mathrm{sinh}\left(\frac{\Phi_{-}^{(+)}}{2}\right)\bar{\Psi}_{-}^{(-)}+\b_{3}\,\mathrm{sinh}\left(\frac{\Phi_{+}^{(-)}}{2}\right)\Psi_{+}^{(+)}\right]\nonumber\\
&&-\frac{2\b_2}{\b_3}\mathrm{sinh}(\varphi_{1}-\varphi_{2})-\frac{2\b_3}{\b_2}\mathrm{sinh}(\varphi_{1}+\varphi_{2})
\end{eqnarray}

\begin{eqnarray}
\partial_x \varphi_2 - \partial_t \varphi_1&=&\frac{i}{2\sqrt{2}}\zeta_{1}^{-}\left[-\b_{2}\,
\mathrm{sinh}\left(\frac{\Phi_{-}^{(-)}}{2}\right)\bar{\Psi}_{-}^{(+)}+\b_{3}\,\mathrm{sinh}\left(\frac{\Phi_{+}^{(+)}}{2}\right)\Psi_{+}^{(-)}\right]\nonumber\\
&&+\frac{i}{2\sqrt{2}}\frac{\b_1}{\b_3}\zeta_{1}^{+}\left[\b_{2}\,
\mathrm{sinh}\left(\frac{\Phi_{-}^{(+)}}{2}\right)\bar{\Psi}_{-}^{(-)}-\b_{3}\,\mathrm{sinh}\left(\frac{\Phi_{+}^{(-)}}{2}\right)\Psi_{+}^{(+)}\right]\nonumber\\
&&-\frac{2\b_2}{\b_3}\mathrm{sinh}(\varphi_{1}-\varphi_{2})+\frac{2\b_3}{\b_2}\mathrm{sinh}(\varphi_{1}+\varphi_{2})
\end{eqnarray}

\begin{eqnarray}\label{vinc1}
\psi_{1}-\psi_{2}=-\frac{i}{2\sqrt{2}}\b_3\zeta_{1}^{-}\mathrm{cosh}\left(\frac{\Phi_{+}^{(+)}}{2}\right)-\frac{i}{2\sqrt{2}}\b_1\zeta_{1}^{+}\mathrm{cosh}\left(\frac{\Phi_{+}^{(-)}}{2}\right)
\end{eqnarray}

\begin{eqnarray}\label{vinc2}
\chi_{1}-\chi_{2}=-\frac{i}{2\sqrt{2}}\b_3\zeta_{1}^{-}\mathrm{cosh}\left(\frac{\Phi_{+}^{(+)}}{2}\right)+\frac{i}{2\sqrt{2}}\b_1\zeta_{1}^{+}\mathrm{cosh}\left(\frac{\Phi_{+}^{(-)}}{2}\right)
\end{eqnarray}

\begin{eqnarray}\label{vinc3}
\bar{\psi}_{1}+\bar{\psi}_{2}=-\frac{i}{2\sqrt{2}}\b_2\zeta_{1}^{-}\mathrm{cosh}\left(\frac{\Phi_{-}^{(-)}}{2}\right)-\frac{i}{2\sqrt{2}}\b_1\frac{\b_2}{\b_3}\zeta_{1}^{+}\mathrm{cosh}\left(\frac{\Phi_{-}^{(+)}}{2}\right)
\end{eqnarray}

\begin{eqnarray}\label{vinc4}
\bar{\chi}_{1}+\bar{\chi}_{2}=\frac{i}{2\sqrt{2}}\b_2\zeta_{1}^{-}\mathrm{cosh}\left(\frac{\Phi_{-}^{(-)}}{2}\right)-\frac{i}{2\sqrt{2}}\b_1\frac{\b_2}{\b_3}\zeta_{1}^{+}\mathrm{cosh}\left(\frac{\Phi_{-}^{(+)}}{2}\right)
\end{eqnarray}

\begin{eqnarray}
\partial_{t}\zeta_{1}^{+}=-i 2\sqrt{2}\frac{\b_3}{\b_1
\b_2}\mathrm{cosh}\left(\frac{\Phi_{+}^{(+)}}{2}\right)\Psi_{+}^{(-)}+i
\frac{2\sqrt{2}}{\b_1}\mathrm{cosh}\left(\frac{\Phi_{-}^{(-)}}{2}\right)\bar{\Psi}_{-}^{(+)}
\end{eqnarray}

\begin{eqnarray}
\partial_{t}\zeta_{1}^{-}=-i \frac{2\sqrt{2}}{\b_2}\mathrm{cosh}\left(\frac{\Phi_{+}^{(-)}}{2}\right)\Psi_{+}^{(+)}+i
\frac{2\sqrt{2}}{\b_3}\mathrm{cosh}\left(\frac{\Phi_{-}^{(+)}}{2}\right)\bar{\Psi}_{-}^{(-)}\label{zeta2}
\label{3.31}
\end{eqnarray}

\begin{eqnarray}
\partial_{x}\zeta_{1}^{+}&=&-i 2\sqrt{2}\frac{\b_3}{\b_1
\b_2}\mathrm{cosh}\left(\frac{\Phi_{+}^{(+)}}{2}\right)\Psi_{+}^{(-)}-i
\frac{2\sqrt{2}}{\b_1}\mathrm{cosh}\left(\frac{\Phi_{-}^{(-)}}{2}\right)\bar{\Psi}_{-}^{(+)}
\label{zet1}
\er
\br
\partial_{x}\zeta_{1}^{-}&=&-i \frac{2\sqrt{2}}{\b_2}\mathrm{cosh}\left(\frac{\Phi_{+}^{(-)}}{2}\right)\Psi_{+}^{(+)}-i
\frac{2\sqrt{2}}{\b_3}\mathrm{cosh}\left(\frac{\Phi_{-}^{(+)}}{2}\right)\bar{\Psi}_{-}^{(-)}
\label{zet2}
\end{eqnarray}

Notice that the above Backlund transformation can be re-obtained from the equations of motion (\ref{6}) with $B$ given by (\ref{15}), (\ref{2.15}) and (\ref{16}){\footnote{The compatibility of eqn. (\ref{6}) with (\ref{3.31}) require $\b_1 \b_2 = -8$}}, that is 
\br
B&=&\frac{i}{\sqrt{2}}\zeta_{1}^{-}\left[\b_{2}\,
\mathrm{cosh}\left(\frac{\Phi_{-}^{(-)}}{2}\right)\bar{\Psi}_{-}^{(+)}-\b_{3}\,\mathrm{cosh}\left(\frac{\Phi_{+}^{(+)}}{2}\right)\Psi_{+}^{(-)}\right]\nonumber\\
&&+\frac{i}{\sqrt{2}}\frac{\b_1}{\b_3}\zeta_{1}^{+}\left[\b_{2}\,
\mathrm{cosh}\left(\frac{\Phi_{-}^{(+)}}{2}\right)\bar{\Psi}_{-}^{(-)}-\b_{3}\,\mathrm{cosh}\left(\frac{\Phi_{+}^{(-)}}{2}\right)\Psi_{+}^{(+)}\right]\nonumber\\
&&+\frac{2\b_2}{\b_3}\mathrm{cosh}(\phi_{1}-\phi_{2})+\frac{2\b_3}{\b_2}\mathrm{cosh}(\phi_{1}+\phi_{2})\nonumber\\
&&-\frac{2\b_2}{\b_3}\mathrm{cosh}(\varphi_{1}-\varphi_{2})-\frac{2\b_3}{\b_2}\mathrm{cosh}(\varphi_{1}+\varphi_{2})
\er
The above Backlund equations  are also invariant under $N=2$ supersymmetry transformation (\ref{supe}).
%\br
%\d (\phi_p \pm \varphi_p) = 2 (\psi_p \pm \chi_p) \eps_{\pm},  \quad \d (\psi_p \pm \chi_p) = - \pa_x (\phi_p \pm \varphi_p) \eps_{\pm},\quad 
%\d  (\bar \psi_p \pm \bar \chi_p) = 2 \sinh (\phi_p \pm \varphi_p) \eps_{\pm}.
% \label{2}
%\nonu
%\er
%%%%%%%%%%%%%%%%%%%%%%%%%%%%%%%%%%%%%%%%
In order to obtain the one-soliton solution let us consider $\phi_2 = \varphi_2 = \psi_2 = \chi_2 \cdots = 0$.  
Since the one-soliton solution contains only one Grassmann
parameter the product of any two Fermi fields vanish and hence  the Backlund equations reduce to a set of two decoupled bosonic sinh-Gordon for $\phi_1$ and $\varphi_1$.
In this case we have
\begin{eqnarray}
\partial_{t}\zeta_1^{\pm}=\lambda_{-}(\mathrm{cosh}\phi_1+\mathrm{cosh}\varphi_1)\zeta_1^{\pm} \quad \quad 
\pa_x \zeta_1^{\pm}=-\lambda_{+}(\mathrm{cosh}\phi_1+\mathrm{cosh}\varphi_1)\zeta_1^{\pm}
\label{zeta}
\end{eqnarray}
where $ \lambda_{\pm}=\left(\frac{\b_2}{\b_3}\pm\frac{\b_3}{\b_2}\right)$. 
Knowing $\phi_1$ and  $\varphi_1$ we can integrate (\ref{zeta}) to  construct the auxiliary fields $\zeta_{1}^{\pm}$ and hence the fermionic fields $\psi_1, \chi_1, \cdots$.

The solution for  $\phi_1$ and  $\varphi_1$ is then
\begin{eqnarray}
\phi_1 = \varphi_1 = \ln\left(\frac{1+\frac{1}{2}b_1
\rho_1}{1-\frac{1}{2}b_1 \rho_1}\right)\nonumber
\end{eqnarray}
where $b_1$ is an arbitrary constant and 
\begin{eqnarray}
\rho_1 =
e^{2(\gamma_1+\gamma_1^{-1})x+2(\gamma_1-\gamma_1^{-1})t}\nonumber
\end{eqnarray}
The above solution satisfy the equation
\begin{eqnarray}
(\partial_{x}^{2}-\partial_{t}^{2})\phi_1=8\,\mathrm{sinh}(2\phi_1)\nonumber
\end{eqnarray}
Integrating (\ref{zeta}) for  $\zeta_1^{\pm}$ and parametrizing 
\begin{eqnarray}
\b_3 = -\gamma_1 \b_2, \qquad \b_2 = i
2\sqrt{2}\nonumber
\end{eqnarray}
we obtain the following solution
\begin{eqnarray}
\zeta_1^{-}&=&\frac{c_1
\gamma_1\rho_1}{1-\frac{1}{4}b_1^{2}\rho_1^{2}}, \qquad \quad\zeta_1^{+}=\gamma_1 \zeta_1^{-}\nonumber\\
\bar{\psi}_1&=&(1-\mathrm{cosh}\phi_1)\zeta_1^{-}, \qquad
\psi_1=\gamma_1
\bar{\psi}_1\nonumber\\
\bar{\chi}_1&=&-(1+\mathrm{cosh}\phi_1)\zeta_1^{-}, \qquad
\chi_1=\gamma_1 \bar{\chi}_1\nonumber
\end{eqnarray}
where $c_1$ is a Grassmann constant. Using the above equations together with the eqns. of motion for the fermions we find
\begin{eqnarray}
\gamma_1(\partial_x - \partial_t)\bar{\psi}_1 &=&
-4\bar{\psi}_1\mathrm{cosh}^{2}\phi_1+4\bar{\chi}_1\mathrm{sinh}^{2}\phi_1\nonumber\\
\gamma_1(\partial_x - \partial_t)\bar{\chi}_1 &=&
-4\bar{\psi}_1\mathrm{sinh}^{2}\phi_1+4\bar{\chi}_1\mathrm{cosh}^{2}\phi_1\nonumber\\
\frac{1}{\gamma_1}(\partial_x + \partial_t)\bar{\psi}_1 &=&
-4\bar{\psi}_1\mathrm{cosh}^{2}\phi_1+4\bar{\chi}_1\mathrm{sinh}^{2}\phi_1\nonumber\\
\frac{1}{\gamma_1}(\partial_x + \partial_t)\bar{\chi}_1 &=&
-4\bar{\psi}_1\mathrm{sinh}^{2}\phi_1+4\bar{\chi}_1\mathrm{cosh}^{2}\phi_1\nonumber
\end{eqnarray}
from where one can verify
\begin{eqnarray}
\gamma_1(\partial_x -
\partial_t)\bar{\psi}_1&=&\frac{1}{\gamma_1}(\partial_x +
\partial_t)\bar{\psi}_1\nonumber\\
\gamma_1(\partial_x -
\partial_t)\bar{\chi}_1&=&\frac{1}{\gamma_1}(\partial_x +
\partial_t)\bar{\chi}_1\nonumber
\end{eqnarray}

\section{Zero Curvature formulation}
In this section we introduce the Lax pair for the $N=2$ super sinh-Gordon model in
 terms of generators of the affine $sl(2,2)$ super Lie algebra (see sect. 3 of 
 ref. \cite{nosso2} for explicit structure of generators). 
  The Lie super algebra $sl(2,2)$ is specified by  the 
 following Bosonic, $\a_1$, $\a_3$  and Fermionic, $\a_2$ simple roots
\br
\a_1 = e_1 - e_2, \quad \a_3 = f_1 - f_2 \quad {\rm and}  \quad \a_2 = e_2 - f_1, \quad \quad e_i \cdot e_j = - f_i \cdot f_j = \d_{ij}
\er
 respectively.

The Lax pair for  the system described by eqn. of motion (\ref{5}) is given by
\begin{eqnarray}
A_x^{(p)}=-\frac{1}{2}\partial_t\phi_p
h_{1}-\frac{1}{2}\partial_t\varphi_p h_{3}+V_{-}^{(p)}\quad \quad 
A_t^{(p)}=-\frac{1}{2}\partial_x\phi_p
h_{1}-\frac{1}{2}\partial_x\varphi_p h_{3}+V_{+}^{(p)}
\end{eqnarray}

where

\begin{eqnarray}
V_{\pm}^{(p)}&=&\left(-e^{-\phi_p}\pm \frac{1}{\lambda}e^{\phi_p}\right)E_{\alpha_1}+(-\lambda
e^{\phi_p}\pm e^{-\phi_p})E_{-\alpha_1}\nonumber\\
&+&(-\lambda
e^{-\varphi_p}\pm e^{\varphi_p})E_{\alpha_3}+\left(-e^{\varphi_p}\pm \frac{1}{\lambda}e^{-\varphi_p}\right)E_{-\alpha_3}+(-\lambda^{1/2}\pm \lambda^{-1/2})I\nonumber\\
&+&i\left[-e^{-\frac{1}{2}(\phi_p-\varphi_p)}(\psi_p+\chi_p)\lambda^{-1/4}
\mp
e^{\frac{1}{2}(\phi_p-\varphi_p)}(\bar{\psi}_p+\bar{\chi}_p)\lambda^{-3/4}\right]E_{\alpha_1
+\alpha_2}\nonumber\\
&+&i\left[-e^{\frac{1}{2}(\phi_p-\varphi_p)}(\psi_p+\chi_p)\lambda^{3/4}
\pm
e^{-\frac{1}{2}(\phi_p-\varphi_p)}(\bar{\psi}_p+\bar{\chi}_p)\lambda^{1/4}\right]E_{-\alpha_1
-\alpha_2}\nonumber\\
&+&i\left[e^{\frac{1}{2}(\phi_p-\varphi_p)}(\psi_p+\chi_p)\lambda^{3/4}
\pm
e^{-\frac{1}{2}(\phi_p-\varphi_p)}(\bar{\psi}_p+\bar{\chi}_p)\lambda^{1/4}\right]E_{\alpha_2
+\alpha_3}\nonumber\\
&+&i\left[e^{-\frac{1}{2}(\phi_p-\varphi_p)}(\psi_p+\chi_p)\lambda^{-1/4}
\mp
e^{\frac{1}{2}(\phi_p-\varphi_p)}(\bar{\psi}_p+\bar{\chi}_p)\lambda^{-3/4}\right]E_{-\alpha_2
-\alpha_3}\nonumber\\
&+&i\left[-e^{-\frac{1}{2}(\phi_p+\varphi_p)}(\psi_p-\chi_p)\lambda^{1/4}
\mp
e^{\frac{1}{2}(\phi_p+\varphi_p)}(\bar{\psi}_p-\bar{\chi}_p)\lambda^{-1/4}\right]E_{\alpha_1
+\alpha_2+\alpha_3}\nonumber\\
&+&i\left[-e^{\frac{1}{2}(\phi_p+\varphi_p)}(\psi_p-\chi_p)\lambda^{1/4}
\pm
e^{-\frac{1}{2}(\phi_p+\varphi_p)}(\bar{\psi}_p-\bar{\chi}_p)\lambda^{-1/4}\right]E_{-\alpha_1
-\alpha_2-\alpha_3}\nonumber\\
&+&i\left[e^{\frac{1}{2}(\phi_p+\varphi_p)}(\psi_p-\chi_p)\lambda^{1/4}
\pm
e^{-\frac{1}{2}(\phi_p+\varphi_p)}(\bar{\psi}_p-\bar{\chi}_p)\lambda^{-1/4}\right]E_{\alpha_2}\nonumber\\
&+&i\left[e^{-\frac{1}{2}(\phi_p+\varphi_p)}(\psi_p-\chi_p)\lambda^{1/4}
\mp
e^{\frac{1}{2}(\phi_p+\varphi_p)}(\bar{\psi}_p-\bar{\chi}_p)\lambda^{-1/4}\right]E_{-\alpha_2}\nonumber
\end{eqnarray}
Here $h_i = \a_i \cdot H, \;\; i=1,2,3$ are the Cartan subalgebra generators, $I = h_1 +2h_2+h_3$  is the identity matrix 
 and $E_{\a}$ denote the  step operators.  
Notice that those step operators associated to a root $\a$ 
containing  the simple root $\a_2$  are fermionic in nature whilst the remaining 
are bosonic.

In order to describe the integrability of the system we  follow \cite{corrigan} and 
 split the space into two overlapping regions, namely, $x\leq b$ and $x\geq a$ with $a<b$.  Inside the overlap region, i.e., $a \leq x \leq b$ 
introduce the following modified Lax pair
 \begin{eqnarray}
\hat{A}_t^{(1)}&=&A_t^{(1)}+\frac{1}{2}\theta(x-a)\Big[(\partial_x\phi_1
- \partial_t\phi_2+\partial_{\phi_1}B)h_1 + (\partial_x\varphi_1 -
\partial_t\varphi_2-\partial_{\varphi_1}B)h_3\nonumber\\
&& +(\psi_1 -
\psi_2+\frac{1}{2}\partial_{\psi_1}B)E_{\alpha_2}+(\chi_1 -
\chi_2-\frac{1}{2}\partial_{\chi_1}B)E_{\alpha_2+\alpha_3}\nonumber\\
&&\left.
+(\partial_t\zeta_1^{+}+\frac{1}{2}\partial_{\zeta_1^{-}}B)E_{\alpha_1+\alpha_2}+(\partial_t\zeta_1^{-}+\frac{1}{2}\partial_{\zeta_1^{+}}B)E_{\alpha_1+\alpha_2+\alpha_3}\right]\nonumber\\
\hat{A}_x^{(1)}&=&\theta(a-x)A_{x}^{(1)}\nonumber\\
\hat{A}_t^{(2)}&=&A_t^{(2)}+\frac{1}{2}\theta(b-x)\Big[(\partial_x\phi_2
- \partial_t\phi_1-\partial_{\phi_2}B)h_1 + (\partial_x\varphi_2 -
\partial_t\varphi_1+\partial_{\varphi_2}B)h_3\nonumber\\
&& +(\bar{\psi}_1 +
\bar{\psi}_2-\frac{1}{2}\partial_{\bar{\psi}_1}B)E_{-\alpha_2}+(\bar{\chi}_1
+
\bar{\chi}_2+\frac{1}{2}\partial_{\bar{\chi}_1}B)E_{-\alpha_2-\alpha_3}\nonumber\\
&&\left.
+(\partial_t\zeta_1^{+}+\frac{1}{2}\partial_{\zeta_1^{-}}B)E_{-\alpha_1-\alpha_2}+(\partial_t\zeta_1^{-}+\frac{1}{2}\partial_{\zeta_1^{+}}B)E_{-\alpha_1-\alpha_2-\alpha_3}\right]\nonumber\\
\hat{A}_x^{(2)}&=&\theta(x-b)A_{x}^{(2)}\nonumber
\end{eqnarray}
Within the overlap region the Lax pair denoted by suffices $p=1,2$ are related by gauge transformation,
\begin{eqnarray}
\partial_t K = K\hat{A}_t^{(2)}-\hat{A}_{t}^{(1)}K. \nonumber
\end{eqnarray}
Decomposing $K$ into
\begin{eqnarray}
K=e^{\frac{1}{2}\phi_2 h_1+\frac{1}{2}\varphi_2
h_3}\bar{K}e^{-\frac{1}{2}\phi_1 h_1-\frac{1}{2}\varphi_1
h_3}\label{k}
\end{eqnarray}
we have
\begin{eqnarray}
(\partial_{\phi_1}B h_1 -\partial_{\varphi_1}B
h_3)\bar{K}+\bar{K}(\partial_{\phi_2}B h_1-\partial_{\varphi_2}B
h_3)=2\bar{K}M-2N\bar{K}-2\partial_t\bar{K}\nonumber
\end{eqnarray}
where
\begin{eqnarray}
M&=&e^{-\frac{1}{2}\phi_1 h_1-\frac{1}{2}\varphi_1
h_3}V_+^{(2)}e^{\frac{1}{2}\phi_1 h_1+\frac{1}{2}\varphi_1
h_3}\nonumber\\
N&=&e^{-\frac{1}{2}\phi_2 h_1-\frac{1}{2}\varphi_2
h_3}V_+^{(1)}e^{\frac{1}{2}\phi_2 h_1+\frac{1}{2}\varphi_2
h_3}\label{mn}
\end{eqnarray}
which are explicitly displayed in the appendix.  
The solution for $\bar K$ is then given in the closed form 
\begin{eqnarray}
\bar{K}&=& C I-\frac{\b_3}{\b_2} C (\lambda^{-1}E_{\a_1}+E_{-\alpha_1}+E_{\alpha_3}+\lambda^{-1}E_{-\alpha_3})\nonumber\\
&&-\frac{C}{\b_2}2\sqrt{2}\lambda^{-3/4}
\zeta_1^{+}(E_{\alpha_1+\alpha_2}-\lambda E_{-\alpha_1
-\alpha_2}-\lambda
E_{\alpha_2+\alpha_3}+E_{-\alpha_2-\alpha_3})\nonumber\\
&&+\frac{C}{2\sqrt{2}}\b_3
\lambda^{-1/4}\zeta_1^{-}(-E_{\alpha_2}+E_{-\alpha_2}+E_{\alpha_1+\alpha_2+\alpha_3}-E_{-\alpha_1-\alpha_2-\alpha_3})\label{k}
\end{eqnarray}
and $ C$ is an arbitrary  constant.  The existence of the gauge transformation  (\ref{k}) provides a generating function for an infinite set of constants of motion (see \cite{bowcock2}) strongly indicating  the integrability of the system.

The existence of Backlund transformations for bosonic and fermionic  systems provide  an interesting class of integrable  models whose mathematical structure deserves further investigation.  For instance its bi-hamiltonian properties (see for instance\cite{magri}). 

The study   of the  quantum  bosonic sinh-Gordon model with defects was  explored in \cite{bowcock3}, \cite{zamb}. The extension     for the  
$N=1$ and $N=2$  super sinh-Gordon model with jump defects is also  of interest. 

These problems are under investigation.

%%%%%%%%%%%%%%%%%%%%%%%%%%%%%%%%%%%%%%%%%%%%%%%%%%%%%%%%%%%%%%%%%%%%%%%%%%%%%%%%%%%%%%%%%%%%%%%%%%%

\section{Appendix A - Backlund Transformation for $N=2$ Super sinh-Gordon}
Consider  the fermionic superfields ${\mathcal{F}}_{1}$, ${\mathcal{F}}_{2}$ and 
${\mathcal{G}}_{1}$,  ${\mathcal{G}}_{2}$ introduced in (\ref{f1}) - (\ref{f2}) and  in (\ref{g1}) - (\ref{g2}) respectively written as 
\begin{eqnarray}
{\mathcal{F}}_{1}=D_{+}(\Xi_{1}^{+}), \qquad
{\mathcal{F}}_{2}=\bar{D}_{+}(\Xi_{2}^{+})\qquad
{\mathcal{G}}_{1}=D_{-}(\Xi_{1}^{-}), \qquad
{\mathcal{G}}_{2}=\bar{D}_{-}(\Xi_{2}^{-})\nonumber
\end{eqnarray}
where
\begin{eqnarray}
\Xi_{1}^{\pm}&=&q_{1}^{\pm}(z^{\pm},\bar{z}^{\pm})+\theta^{\pm}\zeta_{1}^{\pm}(z^{\pm},\bar{z}^{\pm})+\bar{\theta}^{\pm}\zeta_{2}^{\pm}(z^{\pm},\bar{z}^{\pm})+\theta^{\pm}\bar{\theta}^{\pm}q_{2}^{\pm}(z^{\pm},\bar{z}^{\pm})\nonumber\\
\Xi_{2}^{\pm}&=&p_{1}^{\pm}(z^{\pm},\bar{z}^{\pm})+\theta^{\pm}\xi_{1}^{\pm}(z^{\pm},\bar{z}^{\pm})+\bar{\theta}^{\pm}\xi_{2}^{\pm}(z^{\pm},\bar{z}^{\pm})+\theta^{\pm}\bar{\theta}^{\pm}p_{2}^{\pm}(z^{\pm},\bar{z}^{\pm})\nonumber
\end{eqnarray}
are chiral superfields.  In components  we find
\begin{eqnarray}
\bar{D}_{+}{\mathcal{F}}_{1}&=&\frac{2g}{\b_{1}}{\mathrm{sin}}\left(\frac{\Upsilon_{1}^{-}-\Upsilon_{2}^{-}}{2}\right)\nonumber\\
&&\Downarrow\nonumber\\
q_{2}^{+}=\frac{2g}{\b_{1}}{\mathrm{sin}}\left(\frac{\eta_{-}^{(-)}}{2}\right),\quad \quad
\partial_{\bar{z}}\zeta_{1}^{+}&=&\frac{g}{\b_{1}}{\mathrm{cos}}\left(\frac{\eta_{-}^{(-)}}{2}\right)\bar{\psi}_{-}^{(+)},\quad \quad
\partial_{z}\zeta_{2}^{+}=-\frac{g}{\b_{1}}{\mathrm{cos}}\left(\frac{\eta_{-}^{(-)}}{2}\right)\psi_{-}^{(+)},\nonumber\\
\partial_{\bar{z}}\partial_{z}q_{1}^{+}=-\frac{g}{\b_{1}}{\mathrm{cos}}\left(\frac{\eta_{-}^{(-)}}{2}\right)F_{-}^{(-)}&-&\frac{g}{2\b_{1}}{\mathrm{sin}}\left(\frac{\eta_{-}^{(-)}}{2}\right)\psi_{-}^{(+)}\bar{\psi}_{-}^{(+)}\label{q1+} \label{star}
\end{eqnarray}
where we denote  
$\eta_{\pm}^{(-)}=\eta_{1}^{-}\pm \eta_{2}^{-}$,
$\eta_{\pm}^{(+)}=\eta_{1}^{+}\pm \eta_{2}^{+}$, similarly for the other fields.

\begin{eqnarray}
D_{+}{\mathcal{F}}_{2}&=&-\frac{2g}{\b_{2}}{\mathrm{sin}}\left(\frac{\Upsilon_{1}^{-}+\Upsilon_{2}^{-}}{2}\right)\nonumber\\
&&\Downarrow\nonumber\\
p_{2}^{+}=\frac{2g}{\b_{2}}{\mathrm{sin}}\left(\frac{\eta_{+}^{(-)}}{2}\right),\quad \quad 
\partial_{\bar{z}}\xi_{1}^{+}&=&\frac{g}{\b_{2}}{\mathrm{cos}}\left(\frac{\eta_{+}^{(-)}}{2}\right)\bar{\psi}_{+}^{(+)}, \quad \quad 
\partial_{z}\xi_{2}^{+}=-\frac{g}{\b_{2}}{\mathrm{cos}}\left(\frac{\eta_{+}^{(-)}}{2}\right)\psi_{+}^{(+)}\nonumber\\
\partial_{\bar{z}}\partial_{z}p_{1}^{+}=-\frac{g}{\b_{2}}{\mathrm{cos}}\left(\frac{\eta_{+}^{(-)}}{2}\right)F_{+}^{(-)}&-&\frac{g}{2\b_{2}}{\mathrm{sin}}\left(\frac{\eta_{+}^{(-)}}{2}\right)\psi_{+}^{(+)}\bar{\psi}_{+}^{(+)}\nonumber
\end{eqnarray}

\begin{eqnarray}
\bar{D}_{-}{\mathcal{G}}_{1}&=&\frac{2g}{\b_{3}}{\mathrm{sin}}\left(\frac{\Upsilon_{1}^{+}-\Upsilon_{2}^{+}}{2}\right)\nonumber\\
&&\Downarrow\nonumber\\
q_{2}^{-}=\frac{2g}{\b_{3}}{\mathrm{sin}}\left(\frac{\eta_{-}^{(+)}}{2}\right), \quad \quad
\partial_{\bar{z}}\zeta_{1}^{-}&=&\frac{g}{\b_{3}}{\mathrm{cos}}\left(\frac{\eta_{-}^{(+)}}{2}\right)\bar{\psi}_{-}^{(-)}, \quad \quad 
\partial_{z}\zeta_{2}^{-}=-\frac{g}{\b_{3}}{\mathrm{cos}}\left(\frac{\eta_{-}^{(+)}}{2}\right)\psi_{-}^{(-)}\nonumber\\
\partial_{\bar{z}}\partial_{z}q_{1}^{-}=-\frac{g}{\b_{3}}{\mathrm{cos}}\left(\frac{\eta_{-}^{(+)}}{2}\right)F_{-}^{(+)}&-&\frac{g}{2\b_{3}}{\mathrm{sin}}\left(\frac{\eta_{-}^{(+)}}{2}\right)\psi_{-}^{(-)}\bar{\psi}_{-}^{(-)}\nonumber
\end{eqnarray}

\begin{eqnarray}
D_{-}{\mathcal{G}}_{2}&=&-\frac{2g}{\b_{4}}{\mathrm{sin}}\left(\frac{\Upsilon_{1}^{+}+\Upsilon_{2}^{+}}{2}\right)\nonumber\\
&&\Downarrow\nonumber\\
p_{2}^{-}=\frac{2g}{\b_{4}}{\mathrm{sin}}\left(\frac{\eta_{+}^{(+)}}{2}\right)\label{p2-}, \quad \quad
\partial_{\bar{z}}\xi_{1}^{-}&=&\frac{g}{\b_{4}}{\mathrm{cos}}\left(\frac{\eta_{+}^{(+)}}{2}\right)\bar{\psi}_{+}^{(-)}, \quad \quad 
\partial_{z}\xi_{2}^{-}=-\frac{g}{\b_{4}}{\mathrm{cos}}\left(\frac{\eta_{+}^{(+)}}{2}\right)\psi_{+}^{(-)}\nonumber \\
\partial_{\bar{z}}\partial_{z}p_{1}^{-}=-\frac{g}{\b_{4}}{\mathrm{cos}}\left(\frac{\eta_{+}^{(+)}}{2}\right)F_{+}^{(+)}&-&\frac{g}{2\b_{4}}{\mathrm{sin}}\left(\frac{\eta_{+}^{(+)}}{2}\right)\psi_{+}^{(-)}\bar{\psi}_{+}^{(-)}\label{star-star}
\end{eqnarray}

\begin{eqnarray}
D_{+}\Upsilon_{1}^{+}&=&D_{+}\Upsilon_{2}^{+}+\b_{1}{\mathcal{F}}_{1}{\mathrm{cos}}\left(\frac{\Upsilon_{1}^{-}+\Upsilon_{2}^{-}}{2}\right)\nonumber\\
&&\Downarrow\nonumber\\
\psi_{-}^{(-)}=\b_{1}\zeta_{1}^{+}{\mathrm{cos}}\left(\frac{\eta_{+}^{(-)}}{2}\right), \quad \quad 
\partial_{z}\eta_{-}^{(+)}&=&\frac{\b_{1}}{2}{\mathrm{sin}}\left(\frac{\eta_{+}^{(-)}}{2}\right)\zeta_{1}^{+}\psi_{+}^{(+)}+\b_{1}\partial_{z}q_{1}^{+}{\mathrm{cos}}\left(\frac{\eta_{+}^{(-)}}{2}\right)\nonumber\\
F_{-}^{(+)}&=&\b_{1}q_{2}^{+}{\mathrm{cos}}\left(\frac{\eta_{+}^{(-)}}{2}\right)\nonumber
\end{eqnarray}

\begin{eqnarray}
\bar{D}_{+}\Upsilon_{1}^{+}&=&-\bar{D}_{+}\Upsilon_{2}^{+}+\b_{2}{\mathcal{F}}_{2}{\mathrm{cos}}\left(\frac{\Upsilon_{1}^{-}-\Upsilon_{2}^{-}}{2}\right)\nonumber\\
&&\Downarrow\nonumber\\
\bar{\psi}_{+}^{(-)}=\b_{2}\xi_{2}^{+}{\mathrm{cos}}\left(\frac{\eta_{-}^{(-)}}{2}\right), \quad \quad 
\partial_{\bar{z}}\eta_{+}^{(+)}&=&\frac{\b_{2}}{2}{\mathrm{sin}}\left(\frac{\eta_{-}^{(-)}}{2}\right)\xi_{2}^{+}\bar{\psi}_{-}^{(+)}+\b_{2}\partial_{\bar{z}}p_{1}^{+}{\mathrm{cos}}\left(\frac{\eta_{-}^{(-)}}{2}\right)\nonumber\\
F_{+}^{(+)}&=&\b_{2}p_{2}^{+}{\mathrm{cos}}\left(\frac{\eta_{-}^{(-)}}{2}\right)\nonumber
\end{eqnarray}

\begin{eqnarray}
D_{-}\Upsilon_{1}^{-}&=&D_{-}\Upsilon_{2}^{-}+\b_{3}{\mathcal{G}}_{1}{\mathrm{cos}}\left(\frac{\Upsilon_{1}^{+}+\Upsilon_{2}^{+}}{2}\right)\nonumber\\
&&\Downarrow\nonumber \\
\psi_{-}^{(+)}=\b_{3}\zeta_{1}^{-}{\mathrm{cos}}\left(\frac{\eta_{+}^{(+)}}{2}\right), \quad \quad 
\partial_{z}\eta_{-}^{(-)}&=&\frac{\b_{3}}{2}{\mathrm{sin}}\left(\frac{\eta_{+}^{(+)}}{2}\right)\zeta_{1}^{-}\psi_{+}^{(-)}+\b_{3}\partial_{z}q_{1}^{-}{\mathrm{cos}}\left(\frac{\eta_{+}^{(+)}}{2}\right)\nonumber\\
F_{-}^{(-)}&=&\b_{3}q_{2}^{-}{\mathrm{cos}}\left(\frac{\eta_{+}^{(+)}}{2}\right)\nonumber
\end{eqnarray}

\begin{eqnarray}
\bar{D}_{-}\Upsilon_{1}^{-}&=&-\bar{D}_{-}\Upsilon_{2}^{-}+\b_{4}{\mathcal{G}}_{2}{\mathrm{cos}}\left(\frac{\Upsilon_{1}^{+}-\Upsilon_{2}^{+}}{2}\right)\nonumber\\
&&\Downarrow\nonumber\\
\bar{\psi}_{+}^{(+)}=\b_{4}\xi_{2}^{-}{\mathrm{cos}}\left(\frac{\eta_{-}^{(+)}}{2}\right), \quad \quad 
\partial_{\bar{z}}\eta_{+}^{(-)}&=&\frac{\b_{4}}{2}{\mathrm{sin}}\left(\frac{\eta_{-}^{(+)}}{2}\right)\xi_{2}^{-}\bar{\psi}_{-}^{(-)}+\b_{4}\partial_{\bar{z}}p_{1}^{-}{\mathrm{cos}}\left(\frac{\eta_{-}^{(+)}}{2}\right)\nonumber\\
F_{+}^{(-)}&=&\b_{4}p_{2}^{-}{\mathrm{cos}}\left(\frac{\eta_{-}^{(+)}}{2}\right)\nonumber
\end{eqnarray}
Acting  $\bar{D}_{-}$ in eq.  (\ref{f1}) and using eq. (\ref{g2}),
we obtain
\begin{eqnarray}
{\mathcal{F}}_{1}{\mathcal{G}}_{2}=0\nonumber
\end{eqnarray}
which is satisfied when 
\begin{eqnarray}
\zeta_{1}^{+}=\xi_{2}^{-}, \qquad
q_{2}^{+}=\partial_{\bar{z}}p_{1}^{-}, \qquad
p_{2}^{-}=-\partial_{z}q_{1}^{+}, \qquad
\partial_{z}\zeta_{2}^{+}=-\partial_{\bar{z}}\xi_{1}^{-}\label{f1g2}
\end{eqnarray}
Similarly, acting with  $D_{-}$ in  eq. (\ref{f2}) and making use of 
 eq. (\ref{g1}),we obtain 
\begin{eqnarray}
{\mathcal{F}}_{2}{\mathcal{G}}_{1}=0\nonumber
\end{eqnarray}
which is satisfied  when 
\begin{eqnarray}
\zeta_{1}^{-}=\xi_{2}^{+}, \qquad p_{2}^{+}=-\partial_{z}q_{1}^{-},
\qquad q_{2}^{-}=\partial_{\bar{z}}p_{1}^{+}, \qquad
\partial_{z}\zeta_{2}^{-}=-\partial_{\bar{z}}\xi_{1}^{+}\label{f2g1}
\end{eqnarray}
Making use of  eq.  (\ref{f1g2}) and (\ref{star}), we find
\begin{eqnarray}
\partial_{\bar{z}}\partial_{z}q_{1}^{+}=-\partial_{\bar{z}}p_{2}^{-}=-\frac{g}{\b_{1}}{\mathrm{cos}}\left(\frac{\eta_{-}^{(-)}}{2}\right)F_{-}^{(-)}-\frac{g}{2\b_{1}}{\mathrm{sin}}\left(\frac{\eta_{-}^{(-)}}{2}\right)\psi_{-}^{(+)}\bar{\psi}_{-}^{(+)}\label{z1}
\end{eqnarray}
Acting  $\partial_{\bar{z}}$ in  the first eqn. (\ref{star-star}) we obtain
\begin{eqnarray}
\partial_{\bar{z}}p_{2}^{-}&=&\frac{2g}{\b_{4}}\partial_{\bar{z}}\left[{\mathrm{sin}}\left(\frac{\eta_{+}^{(+)}}{2}\right)\right]\label{z2}
\end{eqnarray}
In order to (\ref{z1}) be compatible with  (\ref{z2}), it is necessary 
that 
\begin{eqnarray}
\b_{1}\b_{2}=\b_{3}\b_{4}
\label{beta}
\end{eqnarray}

%%%%%%%%%%%%%%%%%%%%%%%%%%%%%%%%%%%%%%%%%%%%%%%%%%%%%%%%%%%%%%%%%%%%%%%

\section{Appendix - B}

Here we give detailed expression for $M$ and $N$ of eqn. (\ref{mn})  
\begin{eqnarray}
M&=&a_{-}E_{\alpha_1}+b_{+}E_{-\alpha_1}+c_{-}E_{\alpha_3}+d_{+}E_{-\alpha_3}+\lambda_{+}I\nonumber\\
&&-\alpha_{+}^{(2)}E_{\alpha_1+\alpha_2}-\beta_{-}^{(2)}E_{-\alpha_1
-\alpha_2}+\beta_{+}^{(2)}E_{\alpha_2+\alpha_3}+\alpha_{-}^{(2)}E_{-\alpha_2-\alpha_3}\nonumber\\
&&-\gamma_{+}^{(2)}E_{\alpha_1+\alpha_2+\alpha_3}-\delta_{-}^{(2)}E_{-\alpha_1-\alpha_2-\alpha_3}+\delta_{+}^{(2)}E_{\alpha_2}+\gamma_{-}^{(2)}E_{-\alpha_2}\nonumber
\end{eqnarray}

\begin{eqnarray}
N&=&a_{+}E_{\alpha_1}+b_{-}E_{-\alpha_1}+c_{+}E_{\alpha_3}+d_{-}E_{-\alpha_3}+\lambda_{+}I\nonumber\\
&&-\alpha_{+}^{(1)}E_{\alpha_1+\alpha_2}-\beta_{-}^{(1)}E_{-\alpha_1
-\alpha_2}+\beta_{+}^{(1)}E_{\alpha_2+\alpha_3}+\alpha_{-}^{(1)}E_{-\alpha_2-\alpha_3}\nonumber\\
&&-\gamma_{+}^{(1)}E_{\alpha_1+\alpha_2+\alpha_3}-\delta_{-}^{(1)}E_{-\alpha_1-\alpha_2-\alpha_3}+\delta_{+}^{(1)}E_{\alpha_2}+\gamma_{-}^{(1)}E_{-\alpha_2}\nonumber
\end{eqnarray}

where
\begin{eqnarray}
a_{-}&=&(-e^{-\phi_{+}}+\lambda^{-1}e^{-\phi_{-}}), \quad \quad 
b_{+}=(-\lambda e^{\phi_{+}}+e^{\phi_{-}}),\nonumber\\
c_{-}&=&(-\lambda e^{-\varphi_{+}}+e^{-\varphi_{-}}),\quad \quad 
d_{+}=(-e^{\varphi_{+}}+\lambda^{-1}e^{\varphi_{-}}),\nonumber\\
\alpha_{\pm}^{(2)}&=&i\left[e^{-\frac{1}{2}\phi_{+}^{(-)}}(\psi_2+\chi_2)\lambda^{-1/4}\pm
e^{-\frac{1}{2}\phi_{-}^{(-)}}(\bar{\psi}_2+\bar{\chi}_2)\lambda^{-3/4}\right]\nonumber\\
\beta_{\pm}^{(2)}&=&i\left[e^{\frac{1}{2}\phi_{+}^{(-)}}(\psi_2+\chi_2)\lambda^{3/4}\pm
e^{\frac{1}{2}\phi_{-}^{(-)}}(\bar{\psi}_2+\bar{\chi}_2)\lambda^{1/4}\right]\nonumber\\
\gamma_{\pm}^{(2)}&=&i\left[e^{-\frac{1}{2}\phi_{+}^{(+)}}(\psi_2-\chi_2)\lambda^{1/4}\pm
e^{-\frac{1}{2}\phi_{-}^{(+)}}(\bar{\psi}_2-\bar{\chi}_2)\lambda^{-1/4}\right]\nonumber\\
\delta_{\pm}^{(2)}&=&i\left[e^{\frac{1}{2}\phi_{+}^{(+)}}(\psi_2-\chi_2)\lambda^{1/4}\pm
e^{\frac{1}{2}\phi_{-}^{(+)}}(\bar{\psi}_2-\bar{\chi}_2)\lambda^{-1/4}\right]\nonumber\\
\lambda_{\pm}&=&(-\lambda^{1/2}\pm \lambda^{-1/2})\nonumber
\end{eqnarray}
and 
\begin{eqnarray}
a_{+}&=&(-e^{-\phi_{+}}+\lambda^{-1}e^{\phi_{-}}), \qquad 
b_{-}=(-\lambda e^{\phi_{+}}+e^{-\phi_{-}})\nonumber\\
c_{+}&=&(-\lambda e^{-\varphi_{+}}+e^{\varphi_{-}}), \qquad 
d_{-}=(-e^{\varphi_{+}}+\lambda^{-1}e^{-\varphi_{-}}),\nonumber\\
\alpha_{\pm}^{(1)}&=&i\left[e^{-\frac{1}{2}\phi_{+}^{(-)}}(\psi_1+\chi_1)\lambda^{-1/4}\pm
e^{\frac{1}{2}\phi_{-}^{(-)}}(\bar{\psi}_1+\bar{\chi}_1)\lambda^{-3/4}\right]\nonumber\\
\beta_{\pm}^{(1)}&=&i\left[e^{\frac{1}{2}\phi_{+}^{(-)}}(\psi_1+\chi_1)\lambda^{3/4}\pm
e^{-\frac{1}{2}\phi_{-}^{(-)}}(\bar{\psi}_1+\bar{\chi}_1)\lambda^{1/4}\right]\nonumber\\
\gamma_{\pm}^{(1)}&=&i\left[e^{-\frac{1}{2}\phi_{+}^{(+)}}(\psi_1-\chi_1)\lambda^{1/4}\pm
e^{\frac{1}{2}\phi_{-}^{(+)}}(\bar{\psi}_1-\bar{\chi}_1)\lambda^{-1/4}\right]\nonumber\\
\delta_{\pm}^{(1)}&=&i\left[e^{\frac{1}{2}\phi_{+}^{(+)}}(\psi_1-\chi_1)\lambda^{1/4}\pm
e^{-\frac{1}{2}\phi_{-}^{(+)}}(\bar{\psi}_1-\bar{\chi}_1)\lambda^{-1/4}\right]\nonumber
\end{eqnarray}

 \noindent
{\bf Acknowledgements} \\
\vskip .1cm \noindent
{  LHY acknowledges support from Fapesp, JFG and AHZ thank CNPq for partial support.}


\bigskip

 \begin{thebibliography}{99}

\bibitem{mussardo}G. Delfino, G. Mussardo and P. Simonetti,\NPB{432}{1994}{518}, hep-th/9409076


\bibitem{bowcock1} P. Bowcock, E. Corrigan and C. Zambon, \IJMPA{19}{2004}{82}, hep-th/0305022; 

\bibitem{bowcock2}P. Bowcock, E. Corrigan and C. Zambon, {\it J. High Energy Phys.} JHEP {\bf{0401}}(2004)056, hep-th/0401020;



\bibitem{corrigan}E. Corrigan and C. Zambon, \JPA{37}{2004}{L471}, hep-th/0407199
\bibitem{ymai}J.F. Gomes, L. H. Ymai and A.H. Zimerman, \JPA{39}{2006}{7471}, hep-th/0601014


\bibitem{ik}T. Inami and Kanno,  Nucl. Phys. B359 (1990) 201
\bibitem{ku}Kobayashi and Uematsu, Phys. Lett. B264 (1991) 107

\bibitem{nosso1}H. Aratyn, J.F. Gomes and A.H. Zimerman, Nucl. Phys. B676 (2004) 537
\bibitem{nosso2}H. Aratyn, J.F. Gomes, L. H. Ymai and A.H. Zimerman,
``$N=2$ and $N=4$ Supersymmetric mKdV and sinh-Gordon Hierarchies'', hep-th/0409171

\bibitem{chai}M. Chaichian and P. Kulish, \PLB{78}{1978}{413}
\bibitem{nepo}R. Nepomechie, \PLB{509}{2001}{183}
%\bibitem{pla}J.F. Gomes, L. H. Ymai and A.H. Zimerman,  \PLA{359}{2006}{630}, hep-th/0607107
\bibitem{bowcock3}P. Bowcock, E. Corrigan and C. Zambon, {\it J. High Energy Phys.} JHEP {\bf{08}}(2005)023
\bibitem{zamb} E. Corrigan and C. Zambon, ``On Purely Transmitting  Defects in Affine Toda Field Theory'', {\it J. High Energy Phys.} JHEP {\bf 0707},(2007), 001, hep-th 07051066

\bibitem{magri}G. Falqui, F. Magri and M. Pedroni \CMP{197}{1998}{303}
 \end{thebibliography}
\end{document}